\documentclass{article}
\usepackage{spconf,amsmath,graphicx}
\usepackage{subfigure}
\usepackage{flushend}
\newcommand{\beq}{\begin{equation}}
\newcommand{\eeq}{\end{equation}}

\title{A Low-Latency FFT-IFFT Cascade Architecture}
\name{Keshab K. Parhi, Fellow, IEEE}
\address{University of Minnesota at Twin Cities}
\begin{document}
%\ninept
%
\maketitle
\begin{abstract}
This paper addresses the design of a partly-parallel cascaded
FFT-IFFT architecture that does not require any intermediate
buffer. Folding can be used to design partly-parallel
architectures for FFT and IFFT. While many cascaded FFT-IFFT architectures can be designed using various folding sets
for the FFT and the IFFT, for a specified folded FFT architecture, there exists a unique folding set to design the IFFT architecture that does not require an intermediate buffer. Such a
folding set is designed by processing the output of the FFT as
soon as possible (ASAP) in the folded IFFT. Elimination of
the intermediate buffer reduces latency and saves area. The proposed approach is also extended to interleaved processing of multi-channel time-series. The proposed FFT-IFFT cascade architecture saves about $N/2$ memory elements and $N/4$ clock cycles of latency compared to a design with identical folding sets.
For the 2-interleaved FFT-IFFT cascade, the memory and latency savings are, respectively, $N/2$ units and $N/2$ clock cycles, compared to a design with identical folding sets.

\end{abstract}
\begin{keywords}
FFT, IFFT, Folding, Multi-channel FFT
\end{keywords}
\section{Introduction}
\label{sec:intro}

The fast Fourier transform (FFT) algorithm \cite{oppenheim2009discrete} is an important operation in many digital signal processing and machine learning systems. FFT is used for frequency-domain and time-frequency representation of signals, and for extracting features for machine learning classifiers.
FFT is used in applications such as digital communication~\cite{mahdavi2017low,taiwo2017mimo,lin2007design,yang2012mdc}, medical imaging~\cite{sanjeet2023low,zhou2018epileptic}, and convolutional neural network (CNN)~\cite{abtahi2018accelerating,ding2017circnn,lee20202}, in the context of both software and hardware implementations.

\begin{figure}[htbp]
    \centering
    \includegraphics[width=\linewidth]{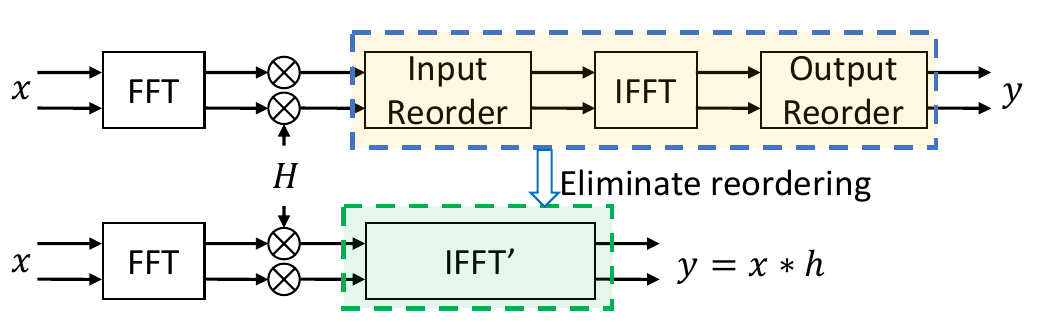}
    \caption{Cascaded FFT-IFFT architecture with and without intermediate buffer.}
    \label{motivation}
\end{figure}

FFT hardware architectures can be classified into two categories: pipelined and parallel architectures~\cite{ayinala2011pipelined,garrido2009pipelined,wang2015mixed,yang2012mdc}, and compact and memory-based FFT architectures~\cite{ayinala2013place, ma2015novel}.
This paper addresses the optimization of the architecture for FFT-IFFT cascade for partly-parallel architectures. In a typical long convolution operation, two FFTs are computed, a pointwise multiplication is carried out, and the IFFT of the product is computed. If the second signal is constant, then the second FFT can be precomputed. It is well known that fully-parallel FFT-IFFT cascades can be realized without using any buffer. This is the first paper to address the architecture for the partly-parallel FFT-IFFT cascade architecture. Partly-parallel, such as 2-parallel and 4-parallel, architectures can be designed by time-multiplexing or folding based on folding sets. If the folding sets are not selected correctly, then an intermediate buffer is needed between the FFT and IFFT hardware blocks, as shown in Fig. \ref{motivation} (top). This increases latency and hardware. This paper shows that by designing a folding set for the IFFT that processes the FFT outputs as soon as possible (ASAP) can eliminate the intermediate buffer requirement, thus reducing latency and area. This is illustrated in Fig. \ref{motivation}(bottom).

The remainder of the paper is organized as follows. Section 2 describes the proposed novel FFT-IFFT cascaded architecture design. Section 3 extends the proposed FFT-IFFT cascade design for processing multi-channel time-series using an interleaving approach. Section 4 compares the area and latency of the proposed approaches with standard designs. 

\section{Cascaded FFT-IFFT Architecture Design}
In this section, we present the limitations and overheads associated with design of a cascaded FFT-IFFT architecture that requires an intermediate buffer. Then we introduce a novel and systematic FFT and IFFT cascaded architecture design via folding transformation~\cite{parhi2007vlsi,parhi1992synthesis} and ASAP scheduling of the FFT outputs at the input of the IFFT.

\subsection{Traditional FFT/IFFT Cascade Architecture}

\begin{figure*}[htbp]
    \centering
    \subfigure[Data-flow graph and schedule for the FFT]{\includegraphics[width=0.310\linewidth]{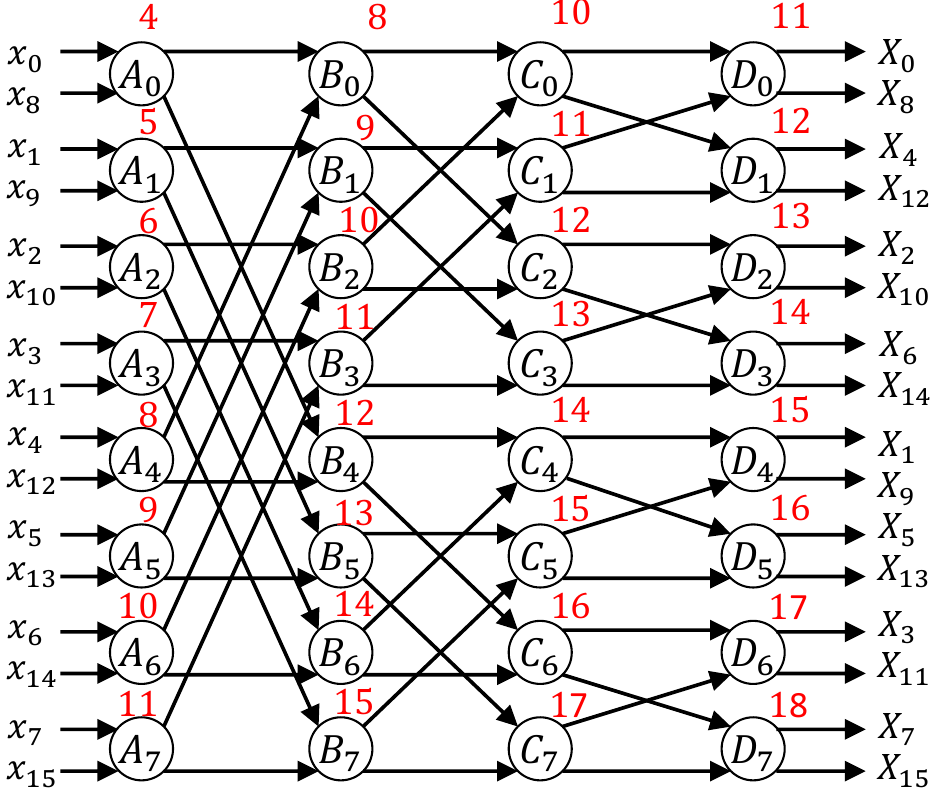}
    % \label{fig:FFT-2-parallel}
    }\hfill
    \subfigure[Data-flow graph and schedule for the traditional IFFT architecture.]{\includegraphics[width=0.333\linewidth]{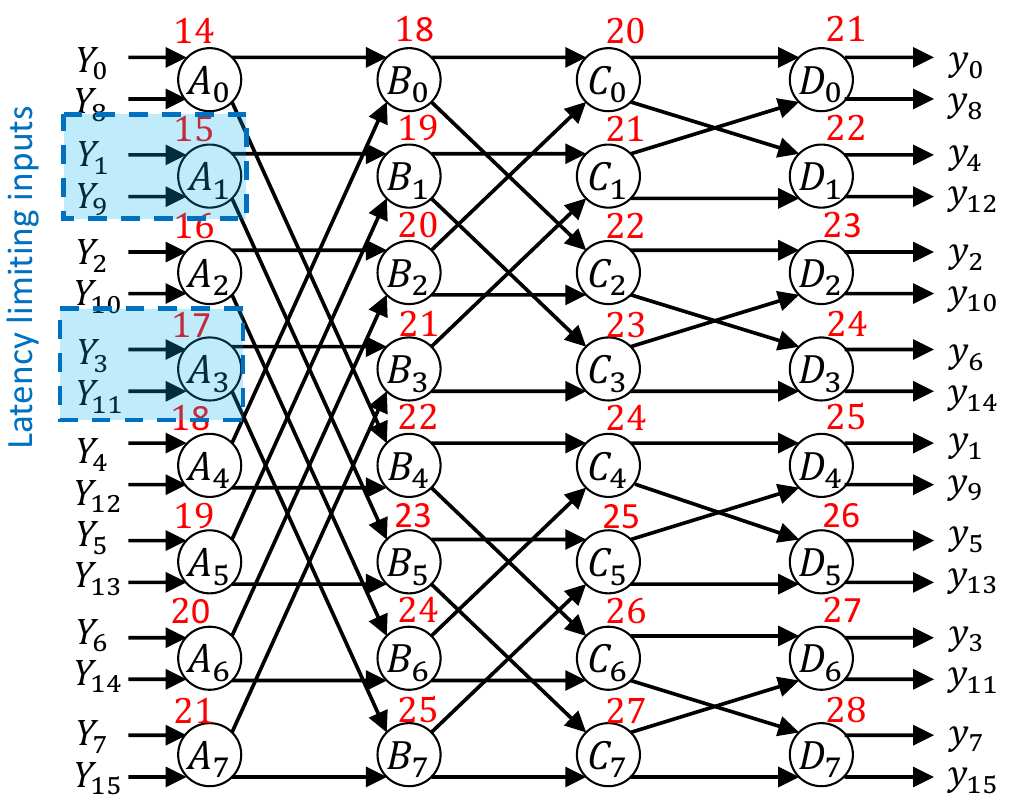}
    % \label{fig:IFFT-2-parallel}
    }\hfill
    \subfigure[Data-flow graph and schedule for the optimized IFFT architecture.]{\includegraphics[width=0.325\linewidth]{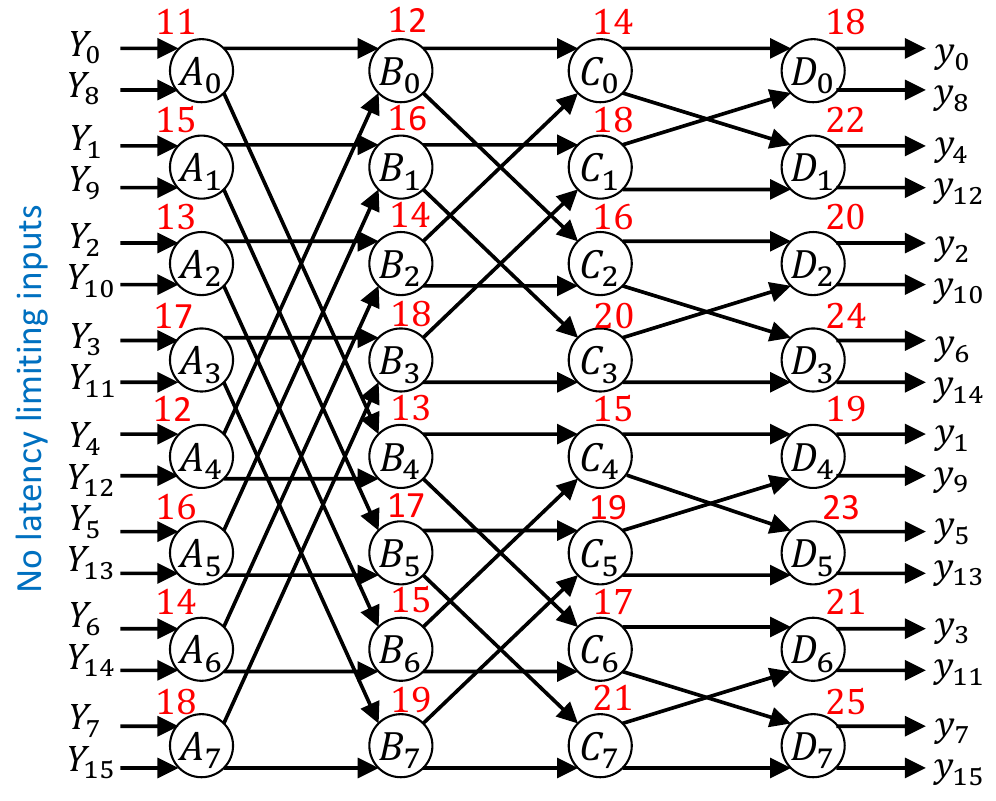}
    }
    \vspace{-12pt}
    \caption{Data-flow graphs for FFT and IFFT with scheduling. Clock cyles are marked in red.}\label{dfg_fft}
    \vspace{-8pt}
\end{figure*}

Consider the flow-graph for a 16-point FFT shown in Fig. \ref{dfg_fft}(a). In a 2-parallel design, two samples are processed in parallel and $N$ samples are processed in $N/2$ clock cycles. Furthermore, the $N/2$ butterflies in every stage are mapped (folded) to one hardware butterfly. This corresponds to a folding factor of $N/2$. Folding sets are not unique, but two types of folding sets are commonly used. In one design, the butterflies are scheduled in consecutive cycles from top to bottom as illustrated in Fig. \ref{dfg_fft}(a) where the clock cycles are marked in red. In another design, the even nodes are scheduled first followed by odd nodes (not described in this paper).

Consider the folding sets for a 16-point FFT for a folding factor of 8 corresponding to the flow-graph in Fig. \ref{dfg_fft}(a):
\begin{equation}
\begin{aligned}
    A = & \left\{ A_4, A_5, A_6, A_7, A_0, A_1, A_2, A_3 \right\}\\
    B = & \left\{ B_0, B_1, B_2, B_3, B_4, B_5, B_6, B_7 \right\}\\
    C = & \left\{ C_6, C_7, C_0, C_1, C_2, C_3, C_4, C_5 \right\}\\
    D = & \left\{ D_5, D_6, D_7, D_0, D_1, D_2, D_3, D_4  \right\}
    \label{FFT-2-parallel}.
\end{aligned}    
\end{equation}

\begin{figure}[htbp]
    \centering
    \includegraphics[width=\linewidth]{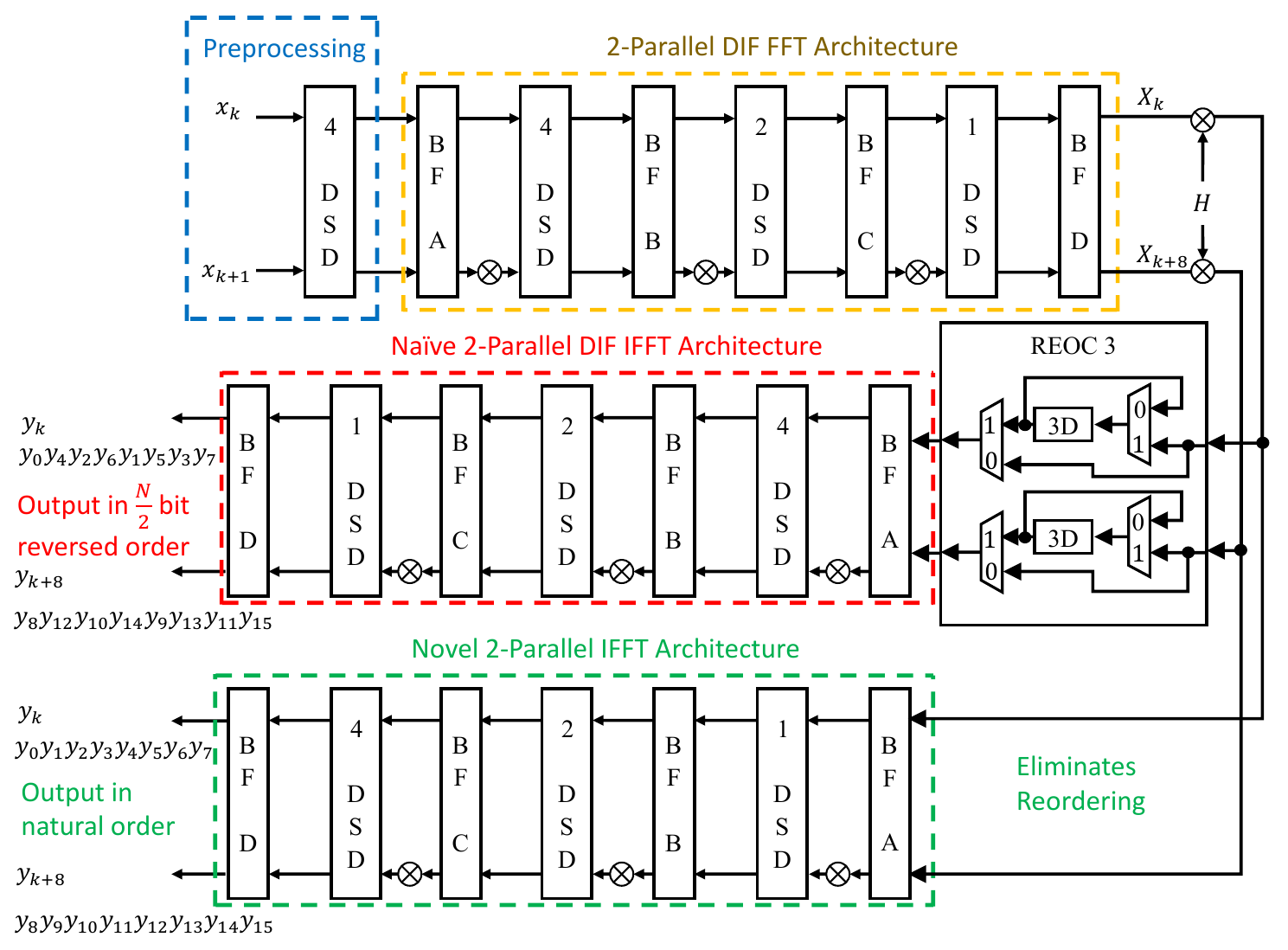}
    \caption{Cascaded 16-Point FFT-IFFT architectures. Top-Middle cascade represents a traditional design. Top-bottom cascade represents the proposed design. }\label{archi_fft}
    % \label{fig:FFT-2-parallel-arch-2}
    \vspace{-12pt}
\end{figure}

In a 2-parallel design, the input $x_8$ is available in clock cycke 4, and the butterfly $A_0$ can be processed as soon as clock cycle 4. Every folding set is associated with a distinct PE from $A$ to $D$. Each element within the set represents the time partition at which the operation is scheduled, given by the clock cycle modulo the folding factor. The timings of these computations establish a direct relation with the folding set. Based on the timings of the butterfly operations, the FFT architecture is depicted in the upper block of Fig. \ref{archi_fft}. Specifically, the delay-switch-delay (DSD) unit is used for the input and output data-flow control of each butterfly block (BF).

A na\"ive approach to implementing the IFFT algorithm is to reuse the same scheduling used for the FFT architecture. Such a folding set is expressed as:
\begin{equation}
\begin{aligned}
    A = & \left\{A_2, A_3, A_4, A_5, A_6, A_7, A_0, A_1 \right\}\\
    B = & \left\{B_6, B_7, B_0, B_1, B_2, B_3, B_4, B_5 \right\}\\
    C = & \left\{C_4, C_5, C_6, C_7, C_0, C_1, C_2, C_3 \right\}\\
    D = & \left\{D_3, D_4, D_5, D_6, D_7, D_0, D_1, D_2  \right\}.
    \label{IFFT-2-parallel}
\end{aligned}    
\end{equation}

The corresponding schedule for the IFFT dataflow graph is illustrated in Fig. \ref{dfg_fft}(b). The hardware IFFT architecture is shown in the middle of Fig. \ref{archi_fft} where the reorder circuit block REOC3 corresponds to an intermediate buffer. The REOC3 buffer comprises two register sets coupled with four multiplexers (MUXs) for data-flow management, and guarantees that the IFFT architecture's input data sequence is aligned correctly. The FFT-IFFT cascade using the top and middle blocks of Fig. \ref{archi_fft} suffers from two drawbacks. First, the intermediate buffer increases area and latency. Second, the outputs of the combined FFT-IFFT cascade using the top and middle blocks are out of order and need to be reordered. These two drawbacks can be overcome by using an ASAP schedule for the IFFT such that the outputs from the FFT can be processed immediately in the IFFT architecture.

\subsection{FFT-IFFT cascade using ASAP Scheduling}
In the proposed systematic approach, the primary design objective lies in eliminating the need for the reordering operation after point-wise multiplication. Consequently, the FFT architecture's output sequence indices must align with the input sequence indices of the IFFT. ASAP scheduling of the FFT outputs requires a specific folding set that is different from that used for the FFT. The IFFT folding set using ASAP scheduling is described by:
\begin{equation}
\begin{aligned}
    A = & \left\{A_5, A_3, A_7, A_0, A_4, A_2, A_6, A_1 \right\}\\
    B = & \left\{B_1, B_5, B_3, B_7, B_0, B_4, B_2, B_6 \right\}\\
    C = & \left\{C_2, C_6, C_1, C_5, C_3, C_7, C_0, C_4 \right\}\\
    D = & \left\{D_3, D_7, D_0, D_4, D_2, D_6, D_1, D_5 \right\}.
    \label{Novel-IFFT-2-parallel}
\end{aligned}    
\end{equation}
In this folding set, the even nodes are scheduled first and the odd nodes are scheduled next; this corresponds to bit-reversed ordering of the nodes.

By leveraging this optimized folding set and seamlessly integrating it within the IFFT's pipelined and parallel architecture, we can effectively eliminate the REOC reordering unit. The proposed FFT-IFFT cascade consists of the top and bottom parts of Fig. \ref{archi_fft}. As elucidated in Fig. \ref{archi_fft} (presented within the green box at the bottom), the required hardware resources are identical to those of the FFT architecture. Furthermore, the output sequence inherently adheres to a natural order.

\section{Interleaved FFT-IFFT Cascade Architecture}
\begin{figure*}[htbp]
    \centering
    \subfigure[Data-flow graph and schedule for an interleaved FFT.]{\includegraphics[width=0.327\linewidth]{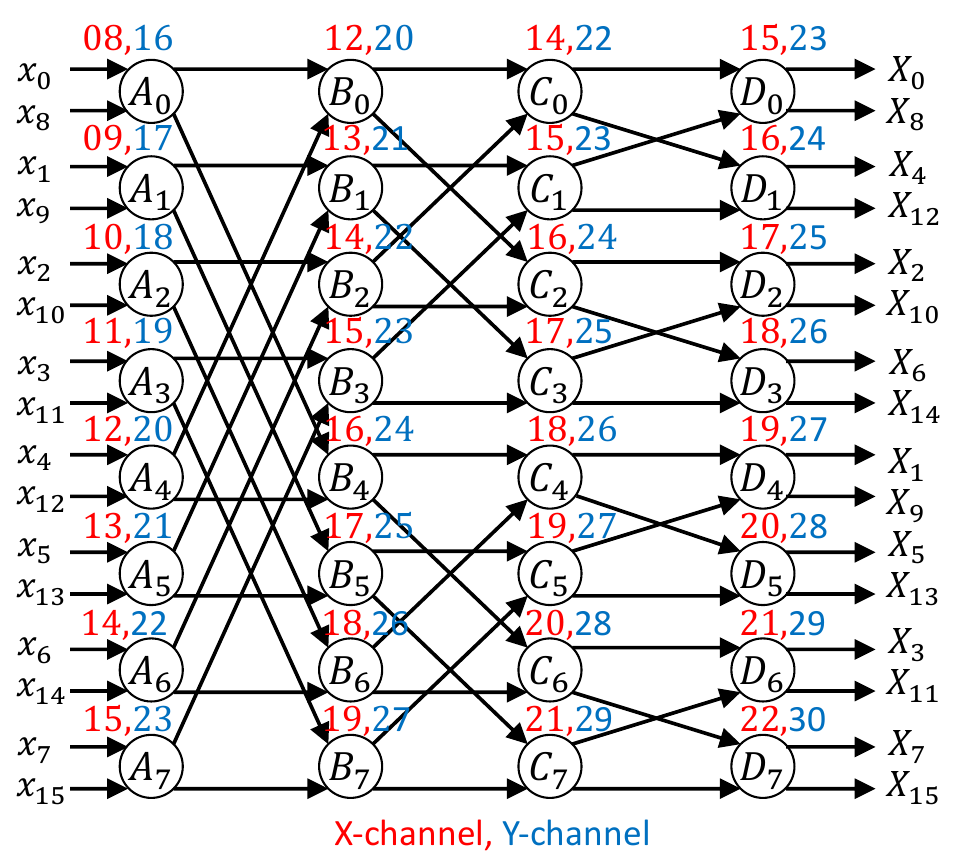}}\hfill
    \subfigure[Data-flow graph and schedule for a traditional interleaved IFFT.]{\includegraphics[width=0.330\linewidth]{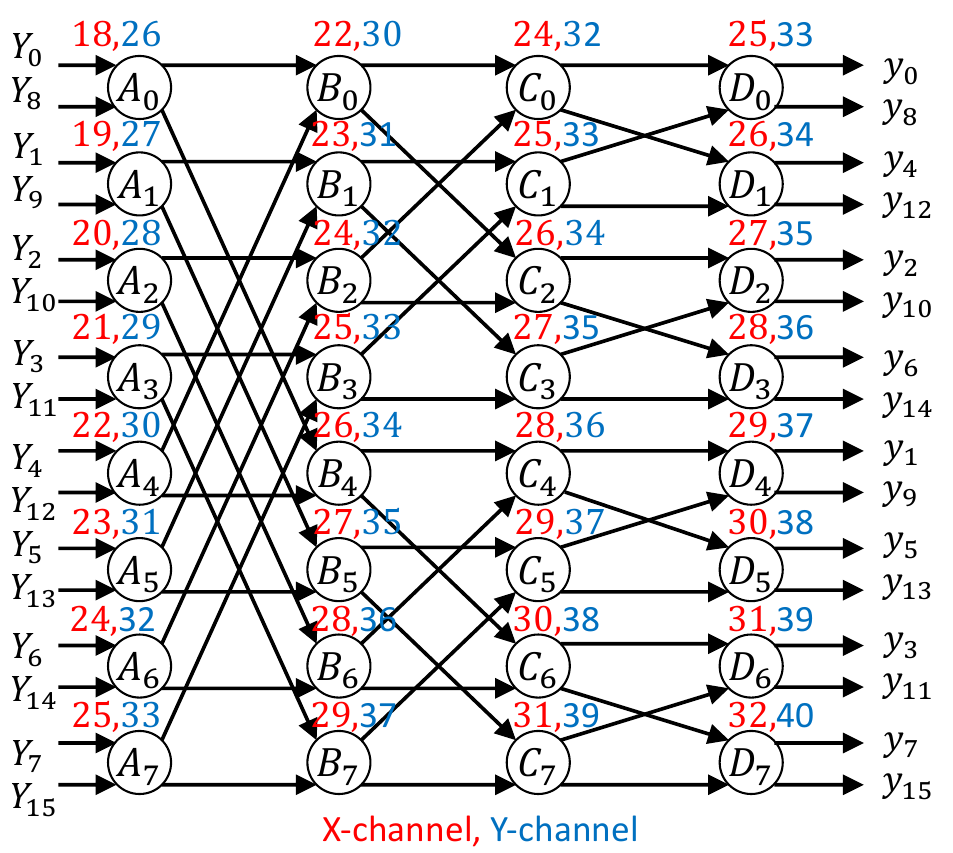}}\hfill
    \subfigure[Data-flow graph and schedule for an optimized interleaved IFFT.]{\includegraphics[width=0.325\linewidth]{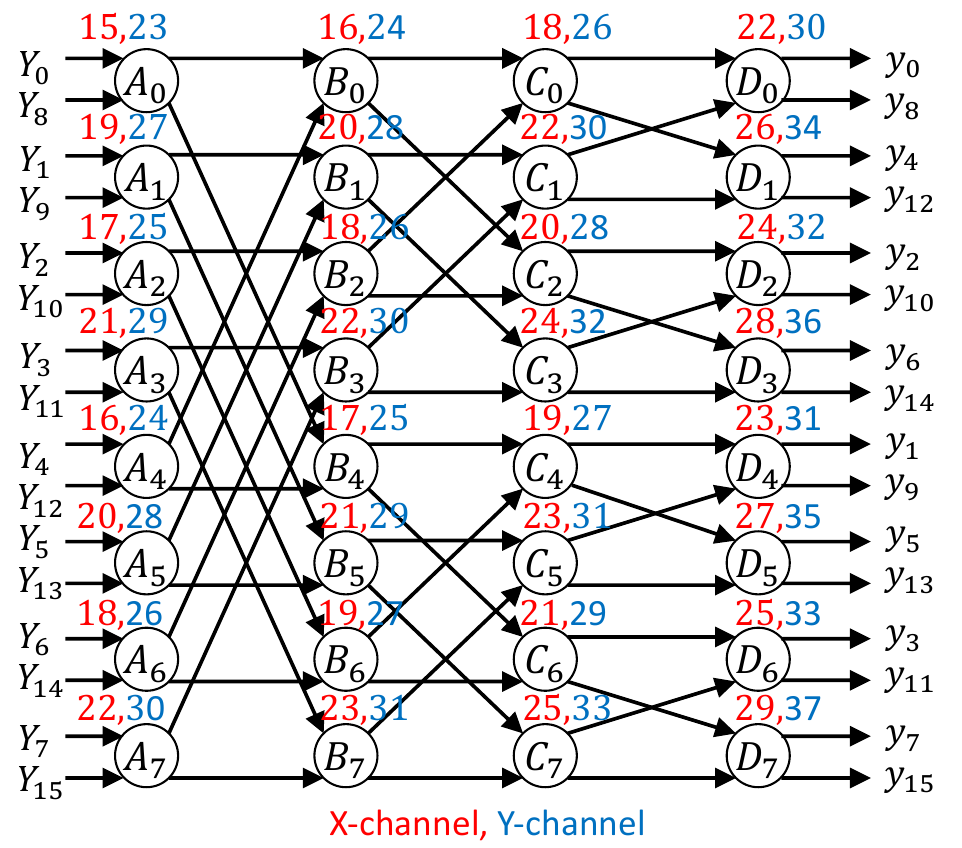}}
    \vspace{-12pt}
    \caption{Data-flow graphs and schedules for Interleaved FFT and IFFT.}\label{dfg_two_channel}
    \vspace{-8pt}
\end{figure*}
The proposed architecture can be extended to a multi-channel framework to design an interleaved FFT and IFFT cascaded architecture. This is realized by adopting the interleaving method~\cite{parhi2007vlsi}, a technique that has found previous applications in numerous domains of digital signal processing and beyond~\cite{parhi2013hierarchical,unnikrishnan2023intergrad,unnikrishnan2022multi}. 

We utilize the folding set delineated in \cite{unnikrishnan2022multi} for the FFT architecture design to begin with our optimization. An example representation of the 16-point FFT computation and its associated folding set is elaborated below:
\begin{equation}
\begin{aligned}
    A = & \big \{A'_0, A'_1, A'_2, A'_3, A'_4, A'_5, A'_6, A'_7,\\
    &  A_0, A_1, A_2, A_3, A_4, A_5, A_6, A_7 \big\}\\
    B = & \big \{B_4, B_5, B_6, B_7, B'_0, B'_1, B'_2, B'_3, \\
    & B'_4, B'_5, B'_6, B'_7, B_0, B_1, B_2, B_3 \big \}\\
    C = & \big \{C_2, C_3, C_4, C_5, C_6, C_7, C'_0, C'_1, \\
    & C'_2, C'_3, C'_4, C'_5, C'_6, C'_7, C_0, C_1 \big \}\\
    D = & \big \{D_1, D_2, D_3, D_4, D_5, D_6, D_7, D'_0, \\
    & D'_1, D'_2, D'_3, D'_4, D'_5, D'_6, D'_7, D_0 \big \}. 
    \label{FFT-2-channel}
\end{aligned}
\end{equation}

In contrast to Eq. (\ref{FFT-2-parallel}), the uniqueness of this folding set is revealed in its composition: it contains two distinct data sequences with an interleaving factor of two. Notably, these sequences denote the X-channel and the Y-channel, respectively. Another feature of this folding set is its inherent efficiency, ensuring a $100\%$ resource utilization across the FFT architecture. The PEs conduct computations for both data sequences in a time-multiplexed and interleaved fashion. This operation is depicted in Fig. \ref{dfg_two_channel}(a), wherein the X-channel and Y-channel execution timings are discernibly represented in colors of red and blue, respectively.

A na\"ive  approach to implementing the IFFT algorithm is to straightforwardly employ the FFT architecture, encompassing the subsequent folding set:
\begin{equation}
\begin{aligned}
    A = & \big \{A'_6, A'_7, A_0, A_1, A_2, A_3, A_4, A_5, \\
    & A_6, A_7, A'_0, A'_1, A'_2, A'_3, A'_4, A'_5 \big \}\\
    B = & \big \{B'_2, B'_3, B'_4, B'_5, B'_6, B'_7, B_0, B_1, \\
    & B_2, B_3, B_4, B_5, B_6, B_7, B'_0, B'_1 \big \}\\
    C = & \big \{C'_0, C'_1, C'_2, C'_3, C'_4, C'_5, C'_6, C'_7, \\
    & C_0, C_1, C_2, C_3, C_4, C_5, C_6, C_7 \big \}\\
    D = & \big \{D_7, D'_0, D'_1, D'_2, D'_3, D'_4, D'_5, D'_6, \\
    & D'_7, D_0, D_1, D_2, D_3, D_4, D_5, D_6 \big \}.
    \label{IFFT-2-channel_unopt}
\end{aligned}
\end{equation}

\begin{figure}[ht]
    \centering
    \includegraphics[width=\linewidth]{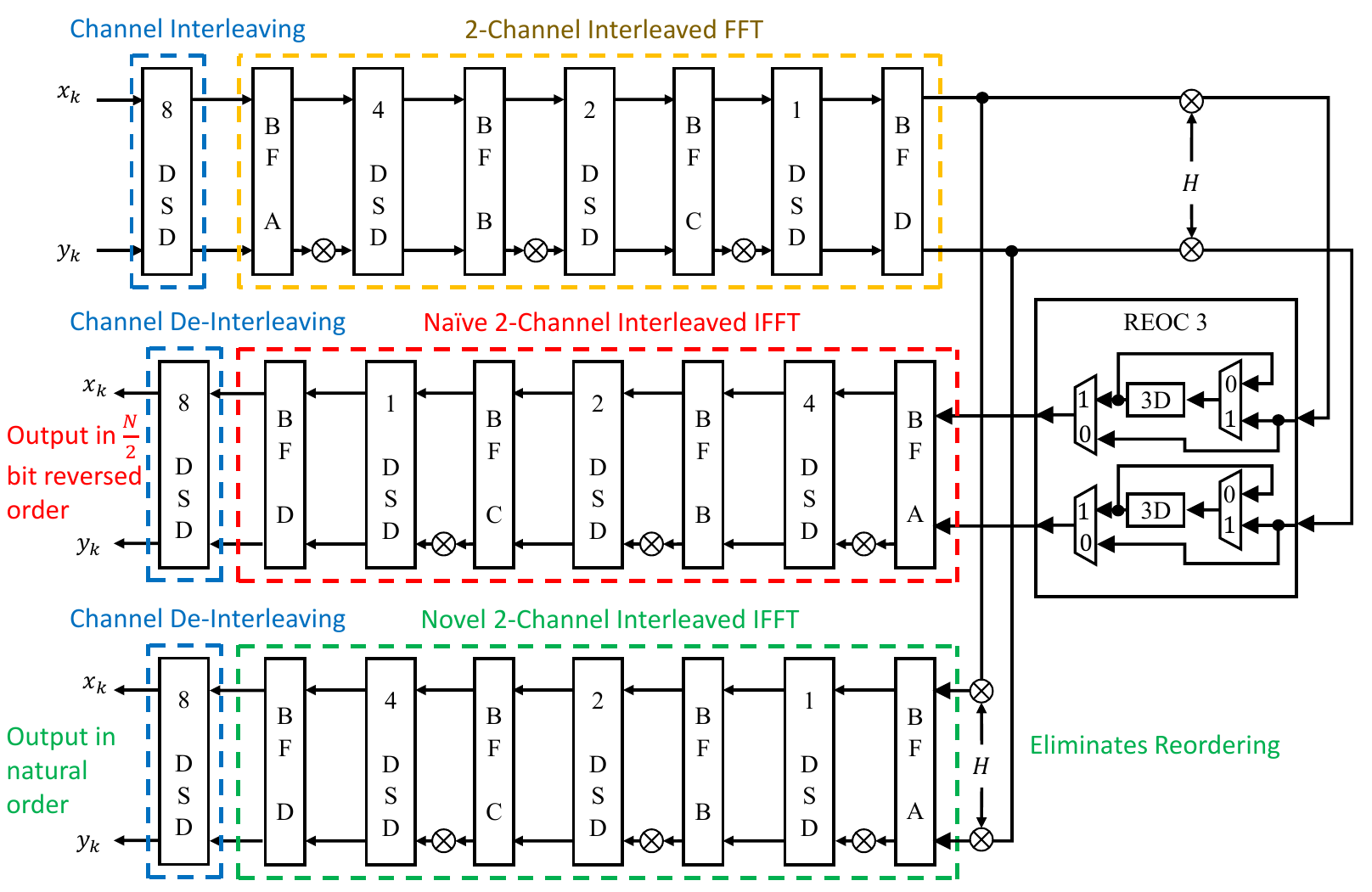}
    \caption{Cascaded interleaved 16-Point FFT-IFFT architectures. Top-Middle cascade represents a traditional design. Top-bottom cascade represents the proposed design.}\label{FFT-2-channel-arch}
\end{figure}

This approach is shown in Fig. \ref{dfg_two_channel}(b) for the data-flow graph and Fig. \ref{FFT-2-channel-arch} (emphasized with a red box) for the architecture, respectively. Notably, while this might appear straightforward, it introduces several inefficiencies in the process. One of the drawbacks is the latency-limiting input, which mandates the use of a REOC unit. Additionally, the architecture consumes two more DSD units engaged in the tasks of de-interleaving and re-interleaving before and after the point-wise multiplication.

\begin{table*}[htbp]
  \centering
  \caption{Performance Analysis and comparison of the proposed designs and the prior works (for $N=1024$)}\label{tb_comp}
\begin{tabular}{|c||c|c|c|c|c|}
\hline
Design &\# BF &\# Memory Elem.&Latency  &\# MUXs &Throughput\\ \hline 
FFT+IFFT \cite{ayinala2011pipelined}   &$\log_2N$ (10)   &$3N-6$ (3066)&$1.5N-3$ (1533) &$4\log_2N+2$ (42)  &2 \\  \hline
\textbf{Proposed I} &$\log_2N$ (10)   &$2.5N-4$ (2556) &$1.25N-2$ (1278) &$4\log_2N-2$ (38)  &2\\  \hline
FFT+IFFT \cite{unnikrishnan2022multi} &$\log_2N$ (10)  &$4.5N-6$ (4602) &$2N-3$ (2045)  &$4\log_2N+4$ (44) &2\\  \hline
\textbf{Proposed II} &$\log_2N$ (10)  &$4N-4$ (4092) &$1.5N-2$ (2046) &$4\log_2N$ (40)  &2\\  \hline
\end{tabular}
\vspace{-12pt}
\end{table*}

Hence, an elaborate exploration into the systematic design for the interleaved FFT and IFFT architecture becomes imperative to remove these unnecessary components for re-ordering. Specifically, the input sequences for both the X-channel and Y-channel in the IFFT are aligned with the output sequences of the FFT architecture. To achieve this, the optimized folding set is presented as follows:
\begin{equation}
\begin{aligned}
   A = & \big \{A_4, A_2, A_6, A_1, A_5, A_3, A_7, A'_0,\\ 
   & A'_4, A'_2, A'_6, A'_1, A'_5, A'_3, A'_7, A_0 \big \}\\
    B = & \big \{B_0, B_4, B_2, B_6, B_1, B_5, B_3, B_7, \\
    & B'_0, B'_4, B'_2, B'_6, B'_1, B'_5, B'_3, B'_7 \big \}\\
    C = & \big \{C'_3, C'_7, C_0, C_4, C_2, C_6, C_1, C_5, \\
    & C_3, C_7, C'_0, C'_4, C'_2, C'_6, C'_1, C'_5 \big \}\\
    D = & \big \{D'_2, D'_6, D'_1, D'_5, D'_3, D'_7, D_0, D_2, \\
    & D_4, D_6, D_1, D_5, D_3, D_7, D'_0, D'_4 \big \}.
    \label{Novel-IFFT-2-channel}
\end{aligned}
\end{equation}

This folding set serves as the basis for deriving the schedule shown in Fig. \ref{dfg_two_channel}(c) and the architecture shown at the bottom of Fig. \ref{FFT-2-channel-arch}. The proposed low-latency and low-area architecture eliminates the need for additional DSD and REOC units, as displayed in Fig. \ref{FFT-2-channel-arch} (enclosed in a green box).

\section{Comparison and Performance Analysis}
In this section, we compare our proposed designs against baseline designs from the prior works, particularly those highlighted in~\cite{ayinala2011pipelined} and~\cite{unnikrishnan2022multi}, with same folding order for FFT and IFFT. Table \ref{tb_comp} presents the timing and area performance metrics for FFT size $N$. 
In particular, timing performance evaluates the first-in first-out latency, as measured in clock cycles, and throughput represents the number of samples processed per clock cycle. Within Table \ref{tb_comp}, the proposed FFT and IFFT cascaded architecture is represented as ``Proposed I", and it is benchmarked against the design from \cite{ayinala2011pipelined}. Concurrently, another proposed $2$-interleaved FFT and IFFT cascaded design, labeled ``Proposed II", is compared against the design in \cite{unnikrishnan2022multi}.

The proposed FFT and IFFT cascaded architecture saves about $N/2$ ($16.67\%$)  memory elements and $N/4$ clock cycles of latency ($16.67\%)$
than its baseline design in~\cite{ayinala2011pipelined}. 
For the interleaved FFT-IFFT cascade, the memory and latency savings are, respectively, $N/2$ units ($11\%$) and $N/2$ clock cycles ($25\%)$, compared to \cite{unnikrishnan2022multi}.

\section{Conclusion}

This paper presents a novel approach for designing partly-parallel cascaded FFT-IFFT architectures where the need for an intermediate buffer is eliminated. Several cascaded architectures can be designed by using different folding sets for the FFT; thus, these architectures are non-unique. The approach can be extended to higher levels of parallelism.
\section{Acknowledgment}
The author is grateful to Dr. Nanda Unnikrishnan and Dr. Weihang Tan for their help in preparation of this paper.

\vfill\pagebreak

\bibliographystyle{IEEEbib}
\bibliography{main}

\end{document}